\documentstyle[12pt]{article}
\begin{document}
\baselineskip=18pt                                            
\pagestyle{plain}
\pagenumbering{arabic}
\vspace{2in}

\today \hspace{\stretch{1}} MUTP 97/17

\vspace*{1cm}

\begin{center}
\baselineskip=24pt
\begin{Large}
{\bf The $C_{2}$ Heat-Kernel Coefficient in the \\ 
Presence of Boundary Discontinuities}
\end{Large}

\vspace*{1cm}

\baselineskip=18pt
J.S. Apps and J.S. Dowker

\vspace*{3mm}

{\it Department of Theoretical Physics\\
The University of Manchester, Manchester, England}

\vspace*{1cm}

Abstract
\end{center}

We consider the heat-kernel on a manifold whose boundary is piecewise smooth.
The set of independent geometrical quantities required to construct an 
expression for the contribution 
of the boundary discontinuities to the $C_{2}$ heat-kernel coefficient 
is derived in the case of a scalar field with Dirichlet and Robin boundary 
conditions. The coefficient is then determined using conformal 
symmetry and evaluation on some specific manifolds. For the Robin case
a perturbation technique is also developed and employed.
The contributions to the smeared heat-kernel 
coefficient and cocycle function are calculated. Some incomplete results for 
spinor fields with mixed conditions are also presented.

\newpage

\section{Introduction}

In the theory of quantum fields on a curved background space, the 
short-time expansion of the heat-kernel has been found to be of paramount 
importance. The coefficients in the expansion are involved in, for example,
statistical field theory and the calculation of vacuum energies. 

Much effort has been expended in finding explicit expressions
for the coefficients but, in view of the rapidly increasing complexity, 
the evaluation of higher and higher coefficients would seem to reach a point 
of diminishing returns. Rather than pursue this path, the present work seeks
to consider a situation where the domain itself is generalized. We do this 
within the context of the $C_{2}$ coefficient which is of particular 
importance in 4 dimensions being linked to the 1-loop quantum effect on the 
Einstein field equations, as well as to the conformal anomaly, of relevance
to quantum cosmology. 

So far this coefficient has been determined in the case where the manifold is 
closed \cite{dewitt} 
or has a smooth boundary \cite{bg},\cite{md}. 
In this paper, we determine $C_{2}$ when the 
boundary is {\it piecewise} smooth. This is an extension of our previous work 
on $C_{3/2}$ in the piecewise smooth case
\cite{ffd}. Our approach is to find the most 
general possible expression for the contribution to $C_{2}$ arising from the 
discontinuity -- a non-trivial problem in itself -- and then to confine it 
using the required conformal symmetries and special-case evaluation. 

This knowledge then enables us to find the expression for the change in the 
effective action 
under a conformal transformation. The latter provides a follow-up to previous 
calculations of the effective action on specific manifolds with smooth 
boundaries \cite{ffd,dowsph1,dowsph2,dowcap,cd}.

The physical motivation for this analysis is that piecewise smooth boundaries,
and manifolds, occur in various idealized situations; for example in 
simplicial approximations to general relativity  
and to quantum gravity, \cite{mandp1,mandp2}. 

Also, internal vertices (conical singularities), which are the periodic
versions of the present structure, appear in the theory of quantum fields on
a black hole background and have been the subject of some activity.

\section{Background}

The quantity of interest is the term $C^{(d)}_{2}$
in the small-$t$
expansion of the integrated heat kernel on a $d$-manifold ${\cal M}$ with 
positive-definite metric
\begin{equation}
\label{hk}
G(t)=\int_{\cal M} d^{d}x \sqrt{g}\: \langle x|e^{-t\Delta}|x \rangle
=\frac{1}{(4\pi t)^{d/2}}\sum_{k=0,1/2,\cdots}C^{(d)}_{k}t^{k}
\end{equation}
for the massless scalar field operator 
\begin{equation}
\Delta=-\nabla^{2}+\xi R
\end{equation}
on ${\cal M}$, where the fields are real, and
obey Dirichlet conditions at the 
boundary $\partial{\cal M}$. $\xi$ is a variable coefficient of coupling to the
Ricci scalar $R$. $\Delta$ is conformally covariant if $\xi$ is equal to
\begin{displaymath}
\xi_{d}=\frac{d-2}{4(d-1)}
\end{displaymath}
Although we consider a massless field for simplicity, our results are easily 
generalised to the non-zero mass case.

In principle, in singular situations, there is the possibility of 
${\rm log}\,t$ terms in the expansion. There are none such in the present
situation and they are henceforth ignored.

The
heat-kernel coefficients can be written as integrals of local geometrical
quantities such as the curvature tensor over ${\cal M}$ and its boundary
$\partial {\cal M}$ if one exists.
A traditional method of determining the
integrand is to write down the most general possible expression and then use
known values of $C^{(d)}_{k}$ on particular manifolds to confine it. This shall
be our approach.

A convenient method of calculating the heat-kernel coefficients in specific
cases where the eigenvalues $\lambda_{i}$
 on ${\cal M}$ are known is via its connection with
the generalized zeta function
\begin{equation}
\zeta_{\cal M}(s)=\sum_{i}\lambda_{i}^{-s}
\end{equation}
where an analytic continuation is involved for ${\cal R}e(s)\leq d/2$.
$\zeta_{\cal M}(s)$ can be written as the Mellin transform of the integrated
heat-kernel, yielding its connection with the $C^{(d)}_{k}$ \cite{minak}:
\begin{equation}
\label{czeta}
C^{(d)}_{k}=\lim_{s\rightarrow 0}\:(4\pi)^{d/2} s\: \Gamma \left( \frac{d}{2}
-k+s \right) \zeta_{\cal M} \left( \frac{d}{2}-k+s \right)
\end{equation}
for $k=0,1/2,1 \cdots$. In this way, using the $\lambda_{i}$ on a particular
manifold to calculate the zeta function gives us the values of the heat-kernel
coefficients on that manifold, and hence information about the general
expressions for the coefficients.

If $\Delta$ and the boundary conditions 
are conformally covariant, then from the change
in the $\lambda_{i}$ brought about by a conformal transformation, it is easily
shown that $\zeta_{\cal M}(0)$, and hence $C^{(d)}_{d/2}$, is conformally
invariant. In other words, $C^{(d)}_{k}$ is conformally invariant in $2k$
dimensions -- this places a large restriction on its form. More 
generally and more quantitatively,
the conformal variation of the heat-kernel coefficients under a transformation
$g_{\mu \nu}\rightarrow 
e^{2\delta \omega (x)}g_{\mu \nu}$ for small $\delta \omega$ is \cite{dowker}
\begin{equation}
\label{noncon}
\delta C^{(d)}_{k}=(d-2k)C^{(d)}_{k}[\delta \omega (x)]
+2C^{(d)}_{k-1}[J \delta \omega (x)]   
\end{equation}
where
\begin{displaymath}
J=(d-1)\left( \xi -\xi_{d}\right) \nabla^{2}
\end{displaymath}    
and the {\it smeared} coefficients $C^{(d)}_{k}[f(x)]$ are defined as
\begin{equation}
\int_{\cal M} d^{d}x \sqrt{g}\: \langle x|e^{-t\Delta}|x \rangle f(x)
=\frac{1}{(4\pi t)^{d/2}}\sum_{k=0,1/2,\cdots}C^{(d)}_{k}[f(x)]t^{k}
\end{equation}

Although our main interest is in determining the effect of boundary
discontinuities on $C^{(d)}_{2}$, we give the result
for the smooth boundary case first. It is found that \cite{bg} 
\begin{eqnarray}
\label{smooth}
C^{(d)}_{2}&=&\frac{1}{360}\int_{\cal M}d^{d}x \sqrt{g}\: 
\bigg[ 2R^{\mu \nu \rho \sigma}R_{\mu \nu \rho \sigma}
-2R^{\mu \nu}R_{\mu \nu} +5(6\xi -1)^{2}R^{2}\bigg] \nonumber \\
& &+\frac{1}{360}\int_{\partial \cal M}d^{d-1}x \sqrt{h}\:
\bigg[ \frac{320}{21}\hbox{tr} \kappa^{3} -\frac{88}{7}\kappa \hbox{tr}
\kappa^{2} +\frac{40}{21}\kappa^{3}
-4R^{\mu \nu}\kappa_{\mu \nu}\nonumber \\
& &-4\kappa R^{\mu \nu}n_{\mu}n_{\nu}+16 R^{\mu \nu \rho 
\sigma}n_{\mu}n_{\rho}\kappa_{\nu \sigma}
+10(1-6\xi)\left(2R\kappa-3\nabla_{n}R\right)\bigg] 
\end{eqnarray}
            
$n^{\mu}$ is the inward-pointing normal to the boundary. With the induced
metric $h_{\mu \nu}=g_{\mu \nu}-n_{\mu}n_{\nu}$ we then define the extrinsic
curvature tensor $\kappa_{\mu \nu}=\kappa_{\nu \mu}=
-h_{\mu}^{\alpha}h_{\nu}^{\beta}\nabla_{\alpha} n_{\beta}$,
with $\kappa=\kappa^{\mu}_{\mu}$, 
$\hbox{tr}\kappa^{2}=\kappa_{\mu\nu}\kappa^{\mu\nu}$. In our
conventions 
$\kappa$ is positive on the surface of a ball.
As we would expect 
from (\ref{noncon}), (\ref{smooth}) is conformally invariant in 4 dimensions 
if $\xi=1/6$. 

We note that the coefficients in (\ref{smooth}) are independent of the 
dimension. This is a general feature, and can be proved by multiplying
${\cal M}$ by a circle. We refer to the book by Gilkey \cite{Gilkey1} 
for general information regarding heat-kernel asymptotics.

\section{Boundary discontinuities}

We shall now consider the situation where $\partial {\cal M}$ is not smooth,
but is made up of a number of boundary parts, $\partial {\cal M}_{i}$. The
intersection ${\cal I}_{ij}$ between two adjacent such parts is a manifold 
of
dimension $d-2$ (i.e. codimension 2). We also require that the ${\cal I}_{ij}$ 
be closed and smooth. 
The hemiball and ball $\times$ interval are examples of this situation.
In general the spectral coefficients will 
now contain  ``edge'' terms involving
integrals over the ${\cal I}_{ij}$, as well as volume and boundary terms.

In addition to the boundary normals $n^{\mu}_{i}$, we define normals 
$\widehat{n}^{(ij)\mu}_{i}$ to ${\cal I}_{ij}$ 
pointing into $\partial {\cal M}_{i}$.
${\cal I}_{ij}$ has induced metric $\gamma^{(ij)}_{\mu \nu}$. 
Then on ${\cal I}_{ij}$, suppressing the
$(ij)$ index for convenience:
\begin{eqnarray}
\label{ng}
& &\gamma_{\mu \nu}=h^{i}_{\mu \nu}-\widehat{n}^{i}_{\mu}
\widehat{n}^{i}_{\nu}   
=h^{j}_{\mu \nu}-\widehat{n}^{j}_{\mu}\widehat{n}^{j}_{\nu} \nonumber \\
& &h^{i}_{\mu \nu}=g_{\mu \nu}-n^{i}_{\mu}n^{i}_{\nu} \nonumber \\ 
& &h^{j}_{\mu \nu}=g_{\mu \nu}-n^{j}_{\mu}n^{j}_{\nu}  
\end{eqnarray}
For a dihedral angle $\theta_{ij}$ between $\partial {\cal M}_{i}$ and
$\partial {\cal M}_{j}$, the normals have the relationship
\begin{eqnarray}
\label{norms}
\widehat{n}^{\mu}_{j}=\widehat{n}^{\mu}_{i}\cos \theta 
+n^{\mu}_{i}\sin \theta \nonumber \\
n^{\mu}_{j}=\widehat{n}^{\mu}_{i}\sin \theta -n^{\mu}_{i}\cos \theta
\end{eqnarray}
We limit ourselves to the case where $\theta$ is a constant on each
intersection.

The codimension-2 contribution to $C^{(d)}_{2}$ will have the form
\begin{displaymath}
\sum_{(ij)}\int_{{\cal I}_{ij}}d^{d-2}x \sqrt{\gamma}\: {\cal S}_{ij}
\end{displaymath}
for some scalar ${\cal S}_{ij}$ depending on the local geometry. Our first
restriction on ${\cal S}_{ij}$ is that 
${\cal S}_{ij}\rightarrow a^{-2}{\cal S}_{ij}$ as
$g_{\mu \nu}\rightarrow a^{2}g_{\mu \nu}$,
where $a$ is a constant, to make $C^{(4)}_{2}$ invariant
under a global conformal (scale) transformation. Expressions with this property
can be split into two types: intrinsic and extrinsic. Intrinsic terms 
must be first
order in the Riemann tensor, contracted with the various metrics and
normals. Extrinsic terms do not involve the Riemann tensor, but contain two
$\nabla_{\mu}$ operators which act on the normals and metrics to produce
extrinsic curvatures. It is easily shown
using dimensional arguments that these are the only types of term
possible.

\vspace*{5mm}

{\bf Intrinsic terms:}

With intrinsic terms, the objects we have to work with are, on each
intersection:
\begin{displaymath}
\begin{array}{l l l l l}
R_{\mu \nu \rho \sigma} & \widehat{R}^{i}_{\mu \nu \rho \sigma} & 
\widehat{R}^{j}_{\mu \nu \rho \sigma} &
\widetilde{R}_{\mu \nu \rho \sigma} & \: \\
\: & \: & \: & \: & \: \\
g_{\mu \nu} & n^{i}_{\mu} &  n^{j}_{\mu} &
\widehat{n}^{i}_{\mu} & \widehat{n}^{j}_{\mu}
\end{array}
\end{displaymath}
$\widehat{R}^{i}_{\mu \nu \rho \sigma}$ and 
$\widetilde{R}_{\mu \nu \rho \sigma}$ are the Riemann curvature tensors of
$\partial {\cal M}_{i}$ and ${\cal I}_{ij}$ respectively, 
formed using the appropriate
induced metrics together with the projected covariant derivative. $h^{i}_{\mu
\nu}$, $h^{j}_{\mu \nu}$ and $\gamma_{\mu \nu}$ are redundant since from the
relations (\ref{ng}) they can be expressed in terms of other quantities.

To construct a general form for ${\cal S}_{ij}$, we have to consider, as far as
intrinsic terms are concerned, every possible independent contraction of the
expressions above which involves a single Riemann tensor. From (\ref{norms}),
we never need to mix up different indices $i$ and $j$ in any particular term,
since any expression which involves a mixture of indices can be written as a
linear combination of terms which are pure in $i$ or $j$. The possible
contractions are
\begin{displaymath}
\begin{array}{l l l}
\rho_{1}=R &
\rho^{i}_{2}=\widehat{R}_{i} &    
\rho_{3}=\widetilde{R} \\
\: & \: & \: \\    
\rho^{i}_{4}=R_{\mu \nu}n^{\mu}_{i}\widehat{n}^{\nu}_{i} &    
\rho^{i}_{5}=R_{\mu \nu}\widehat{n}^{\mu}_{i}\widehat{n}^{\nu}_{i} &
\rho^{i}_{6}=R_{\mu \nu}n^{\mu}_{i}n^{\nu}_{i} \\
\: & \: & \: \\   
\rho^{i}_{7}=\widehat{R}^{i}_{\mu \nu}\widehat{n}^{\mu}_{i}
\widehat{n}^{\nu}_{i} &
\rho^{i}_{8}=R_{\mu \nu \rho \sigma}
n^{\mu}_{i}\widehat{n}^{\nu}_{i}n^{\rho}_{i}\widehat{n}^{\sigma}_{i} & \:
\end{array}
\end{displaymath}
and obviously also $\rho^{j}_{2}$, $\rho^{j}_{4}$ etc.
All others are expressible in terms of the above set using the symmetries of
the Riemann tensors, or vanish since
\begin{displaymath}
\widehat{R}^{i}_{\mu \nu \rho \sigma}n^{\mu}_{i}
=\widetilde{R}_{\mu \nu \rho \sigma}n^{\mu}_{i}
=\widetilde{R}_{\mu \nu \rho \sigma}\widehat{n}^{\mu}_{i}=0
\end{displaymath}

In fact, not all of the above quantities are independent, since 
Gauss' equations
\begin{equation}
\label{gauss}
\widehat{R}^{\mu}_{i\: \nu \rho \sigma}
=\kappa^{\mu}_{i\: \rho}\kappa^{i}_{\nu \sigma}
-\kappa^{\mu}_{i\: \sigma}\kappa^{i}_{\nu \rho}
+h^{\alpha}_{i\: \nu}h^{\beta}_{i\: \rho}h^{\gamma}_{i\: \sigma}
R^{\mu}_{\:\: \alpha \beta \gamma}
\end{equation}
and (defining $\widehat{\kappa}^{i}_{\mu \nu}$ to be the extrinsic curvature of
${\cal I}_{ij}$ with respect to $\partial {\cal M}_{i}$) 
\begin{equation}
\label{gausshat}
\widetilde{R}^{\mu}_{\:\: \nu \rho \sigma}=\widehat{\kappa}^{\mu}_{i\: \rho}
\widehat{\kappa}^{i}_{\nu \sigma}
-\widehat{\kappa}^{\mu}_{i\: \sigma}\widehat{\kappa}^{i}_{\nu \rho}
+\gamma^{\alpha}_{\nu}\gamma^{\beta}_{\rho}\gamma^{\gamma}_{\sigma}
\widehat{R}^{\mu}_{i\: \alpha \beta \gamma}
\end{equation}
provide relations between them, assuming we include the required extrinsic
terms, which will be dealt with
separately. Contracting (\ref{gauss}) and (\ref{gausshat}) 
with $g^{\rho}_{\mu}g^{\nu \sigma}$, and (\ref{gauss}) with
$g^{\rho}_{\mu}\widehat{n}^{\nu}\widehat{n}^{\sigma}$, we get, respectively
\begin{eqnarray*}
& &2\rho^{i}_{6}=\rho_{1}-\rho^{i}_{2}
+\kappa^{2}_{i}-\hbox{tr}\kappa^{2}_{i} \\    
& &2\rho^{i}_{7}=\rho^{i}_{2}-\rho_{3}
+\widehat{\kappa}^{2}_{i}-\hbox{tr}\widehat{\kappa}^{2}_{i} \\    
& &\rho^{i}_{8}=\rho^{i}_{5}-\rho^{i}_{7}
+\kappa_{i} \kappa^{i}_{\mu \nu}\widehat{n}^{\mu}_{i}\widehat{n}^{\nu}_{i}
-\kappa^{\mu}_{i\: \rho}\kappa^{i}_{\mu \nu}\widehat{n}^{\rho}_{i}
\widehat{n}^{\nu}_{i}
\end{eqnarray*}
A further relation can be found using (\ref{norms}):
\begin{displaymath}
\rho^{i}_{5}=\rho^{j}_{6}\hbox{cosec}^{2}\theta
-\rho^{i}_{6}\hbox{cot}^{2}\theta +2\rho^{i}_{4}\hbox{cot}\theta
\end{displaymath}

We choose the $\rho_{1}$ to $\rho_{4}$ as our basis. ${\cal S}_{ij}$ must be
symmetric in $i$ and $j$, so it has the form
\begin{equation}
{\cal S}_{ij}=a_{1}R +a_{2}\left( \widehat{R}_{i}+\widehat{R}_{j} \right)
+a_{3}\widetilde{R} +a_{4}R_{\mu \nu}\left( n^{\mu}_{i}\widehat{n}^{\nu}_{i}+
n^{\mu}_{j}\widehat{n}^{\nu}_{j}\right) +\hbox{extrinsic terms}
\end{equation}
Note that technically {\it one} of the 
$R_{\mu \nu}n^{\mu}_{i}\widehat{n}^{\nu}_{i}$ is still superfluous -- 
we could more
economically replace the last term by $a_{4}R_{\mu \nu}n^{\mu}_{i}n^{\nu}_{j}$.
However we choose not to do this since the expression we have vanishes on an
Einstein space, which will turn out to be convenient.

\vspace*{5mm}

{\bf Extrinsic terms:} 

These terms must contain either two extrinsic curvature tensors, or the
derivative of one. In the former case, we have the combinations
\begin{displaymath}
\begin{array}{l l l}
\kappa^{i}_{\mu \nu}\kappa^{i}_{\rho \sigma} &
\kappa^{i}_{\mu \nu}\kappa^{j}_{\rho \sigma} &
\kappa^{i}_{\mu \nu}\widehat{\kappa}^{i}_{\rho \sigma} \\
\: & \: & \: \\
\kappa^{i}_{\mu \nu}\widehat{\kappa}^{j}_{\rho \sigma} &
\widehat{\kappa}^{i}_{\mu \nu}\widehat{\kappa}^{i}_{\rho \sigma} &
\widehat{\kappa}^{i}_{\mu \nu}\widehat{\kappa}^{j}_{\rho \sigma}
\end{array}
\end{displaymath}
plus the expressions created by swapping $i$ for $j$. The extrinsic curvature
of ${\cal I}_{ij}$ with respect to $\partial {\cal M}_{i}$ is defined as 
$\widehat{\kappa}^{i}_{\mu \nu}=\widehat{\kappa}^{i}_{\nu \mu}=
-\gamma^{\alpha}_{\mu} \gamma^{\beta}_{\nu}
\nabla_{\alpha}\widehat{n}^{i}_{\beta}$. We contract the above with
\begin{displaymath}
\begin{array}{l l l l l}
\gamma^{\mu \nu} & n^{\mu}_{i} & n^{\mu}_{j} & \widehat{n}^{\mu}_{i} &
\widehat{n}^{\mu}_{j} 
\end{array}
\end{displaymath}
It is not necessary to include terms where all the indices of $\kappa^{i}_{\mu
\nu}$ or $\kappa^{j}_{\mu \nu}$ are contracted with $\gamma$, since these can
be rewritten in terms of $\widehat{\kappa}^{i}_{\mu \nu}$ and 
$\widehat{\kappa}^{j}_{\mu \nu}$:
\begin{displaymath}
\gamma^{\mu \alpha}\gamma^{\nu \beta}\kappa^{i}_{\alpha \beta}
=-\gamma^{\mu \alpha}\gamma^{\nu \beta}\nabla_{\alpha}n^{i}_{\beta}
=\widehat{\kappa}^{\mu \nu}_{j}\hbox{cosec}\theta 
-\widehat{\kappa}^{\mu \nu}_{i}\hbox{cot}\theta 
\end{displaymath}
where we have used (\ref{norms}). Additionally, (\ref{norms}) means that we can
choose not to contract a $\kappa_{\mu \nu}$ or $\widehat{\kappa}_{\mu \nu}$ 
with an
$n^{\mu}$ or $\widehat{n}^{\mu}$ of different indices $i$ and $j$.

We are left with the contractions
\begin{displaymath}
\begin{array}{l l}
k^{i}_{1}=\kappa^{i}_{\mu \nu}\kappa^{i}_{\rho \sigma}
\widehat{n}^{\mu}_{i}\widehat{n}^{\nu}_{i}\widehat{n}^{\rho}_{i}
\widehat{n}^{\sigma}_{i} &
k^{i}_{2}=\kappa^{i}_{\mu \nu}\kappa^{i}_{\rho \sigma}
\gamma^{\mu \rho}\widehat{n}^{\nu}_{i}\widehat{n}^{\sigma}_{i} \\
\: & \: \\
k_{3}=\kappa^{i}_{\mu \nu}\kappa^{j}_{\rho \sigma}
\widehat{n}^{\mu}_{i}\widehat{n}^{\nu}_{i}\widehat{n}^{\rho}_{j}
\widehat{n}^{\sigma}_{j} &
k_{4}=\kappa^{i}_{\mu \nu}\kappa^{j}_{\rho \sigma}
\gamma^{\mu \rho}\widehat{n}^{\nu}_{i}\widehat{n}^{\sigma}_{j} \\
\: & \: \\
k_{5}^{i}=\kappa^{i}_{\mu \nu}\widehat{\kappa}^{i}_{\rho
\sigma}\widehat{n}^{\mu}_{i}\widehat{n}^{\nu}_{i}\gamma^{\rho \sigma} &
k_{6}^{i}=\kappa^{i}_{\mu \nu}\widehat{\kappa}^{j}_{\rho
\sigma}\widehat{n}^{\mu}_{i}\widehat{n}^{\nu}_{i}\gamma^{\rho \sigma} \\
\: & \: \\
k_{7}^{i}=\widehat{\kappa}^{i}_{\mu \nu}\widehat{\kappa}^{i}_{\rho
\sigma}\gamma^{\mu \nu}\gamma^{\rho \sigma} &
k_{8}^{i}=\widehat{\kappa}^{i}_{\mu \nu}\widehat{\kappa}^{i}_{\rho
\sigma}\gamma^{\mu \rho}\gamma^{\nu \sigma} \\
\: & \: \\
k_{9}=\widehat{\kappa}^{i}_{\mu \nu}\widehat{\kappa}^{j}_{\rho
\sigma}\gamma^{\mu \nu}\gamma^{\rho \sigma} &
k_{10}=\widehat{\kappa}^{i}_{\mu \nu}\widehat{\kappa}^{j}_{\rho
\sigma}\gamma^{\mu \rho}\gamma^{\nu \sigma} \\
\end{array}
\end{displaymath}                                       
as well as $k^{j}_{1}$, $k^{j}_{2}$ etc. Everything else vanishes, since
\begin{displaymath}
\kappa^{i}_{\mu \nu}n^{\nu}_{i}=\widehat{\kappa}^{i}_{\mu \nu}n^{\nu}_{i}
=\widehat{\kappa}^{i}_{\mu \nu}\widehat{n}^{\nu}_{i}=0
\end{displaymath}
The fact that we have restricted $\theta$ to be constant provides an extra
constraint $\gamma^{\mu \nu}\nabla_{\nu}(n^{\alpha}_{i}n^{j}_{\alpha})=0$,
giving
\begin{displaymath}
\kappa^{j}_{\rho \sigma}\gamma^{\mu \rho}\widehat{n}^{\sigma}_{j}  
=-\kappa^{i}_{\rho \sigma}\gamma^{\mu \rho}\widehat{n}^{\sigma}_{i}
\Rightarrow k_{4}=-k^{i}_{2}=-k^{j}_{2} 
\end{displaymath}  
and we discard $k_{4}$. 

In another possible representation, we could use only 
$\kappa^{i}_{\mu \nu}$ and $\kappa^{j}_{\mu \nu}$, since from (\ref{norms})
\begin{displaymath}
\gamma^{\mu \alpha}\gamma^{\nu \beta}\widehat{\kappa}^{i}_{\alpha \beta}
=-\gamma^{\mu\alpha}\gamma^{\nu\beta}\nabla_{\alpha}n^{i}_{\beta}
=\gamma^{\mu \alpha}\gamma^{\nu \beta} \left(
\kappa_{\alpha \beta}^{j}\hbox{cosec}\theta 
+\kappa_{\alpha \beta}^{i}\hbox{cot}\theta \right)
\end{displaymath}
However, we prefer some less alien expressions, and will use in our
calculations
\begin{eqnarray*}
& &\kappa^{2}_{i}=k^{j}_{7}\hbox{cosec}^{2}\theta
-2k_{9}\hbox{cot}\theta \hbox{cosec}\theta
+k^{i}_{7}\hbox{cot}^{2}\theta
+2\left( k^{i}_{6}\hbox{cosec}\theta -k^{i}_{5}\hbox{cot}\theta
\right) +k^{i}_{1} \\
& &\hbox{tr}\kappa^{2}_{i}=k^{j}_{8}\hbox{cosec}^{2}\theta
-2k_{10}\hbox{cot}\theta \hbox{cosec}\theta
+k^{i}_{8}\hbox{cot}^{2}\theta +k^{i}_{1}+2k^{i}_{2} \\
& &\kappa_{i}\kappa_{j}=-\left( k^{i}_{7}+k^{j}_{7}\right)
\hbox{cosec}\theta \hbox{cot}\theta
+k_{9}\left( \hbox{cosec}^{2}\theta +\hbox{cot}^{2}\theta \right)
+\left( k^{i}_{5}+k^{j}_{5}\right) \hbox{cosec}\theta \\
& &\:\:\:\:\:\:\:\:\:\:\:\:\:\: 
-\left( k^{i}_{6}+k^{j}_{6}\right) \hbox{cot}\theta +k_{3} \\
& &\kappa_{i}\widehat{\kappa}_{i}=k_{9}\hbox{cosec}\theta 
-k^{i}_{7}\hbox{cot}\theta +k^{i}_{5} \\
& &\kappa_{i}\widehat{\kappa}_{j}=k^{j}_{7}\hbox{cosec}\theta
-k_{9}\hbox{cot}\theta +k^{i}_{6} \\
& &\widehat{\kappa}^{2}_{i}=k^{i}_{7} \\
& &\hbox{tr}\widehat{\kappa}^{2}_{i}=k^{i}_{8} \\
& &\widehat{\kappa}_{i}\widehat{\kappa}_{j}=k_{9} \\
& &\hbox{tr}\widehat{\kappa}_{i}\widehat{\kappa}_{j}=k_{10}
\end{eqnarray*}
These are manifestly linearly independent (assuming the $k$'s are). In fact,
since $k^{i}_{2}=k^{j}_{2}$ we could discard, for example,
$\hbox{tr}\kappa^{2}_{i}$ {\bf or} $\hbox{tr}\kappa^{2}_{j}$ but to keep 
$i\leftrightarrow j$ symmetry we do not do this.

\vspace*{5mm}

{\bf Derivative terms:}

It remains to consider terms which contain 
the derivative of the extrinsic curvature. The tensors involved are
\begin{displaymath}
\begin{array}{l l}
\widehat{\nabla}_{\mu}\kappa^{i}_{\nu \rho} & 
\widehat{\nabla}_{\mu}\kappa^{j}_{\nu \rho} \\
\: & \: \\
\widetilde{\nabla}_{\mu}\widehat{\kappa}^{i}_{\nu \rho} &
\widetilde{\nabla}_{\mu}\widehat{\kappa}^{j}_{\nu \rho}
\end{array}
\end{displaymath}
where $\widehat{\nabla}_{\mu}$ and $\widetilde{\nabla}_{\mu}$ are the covariant
derivatives projected onto $\partial {\cal M}_{i}$ and ${\cal I}_{ij}$ 
respectively:
\begin{displaymath}
\widehat{\nabla}_{\mu}\kappa^{i}_{\nu \rho}=
h^{i\: \alpha}_{\mu}h^{i\: \beta}_{\nu}h^{i\: \gamma}_{\rho}
\nabla_{\alpha}\kappa^{i}_{\beta \gamma},\:\:\:
\widetilde{\nabla}_{\mu}\widehat{\kappa}^{i}_{\nu \rho}=
\gamma^{\alpha}_{\mu}\gamma^{\beta}_{\nu}\gamma^{\gamma}_{\rho}
\nabla_{\alpha}\widehat{\kappa}^{i}_{\beta \gamma}
\end{displaymath}   
Any other projections of the covariant derivative give nothing new, e.g.
\begin{displaymath}
h^{\mu}_{i\: \alpha}n^{\nu}_{i}\nabla_{\mu}\kappa^{i}_{\nu \rho}
=\kappa^{i}_{\nu \rho}\kappa^{i\: \nu}_{\alpha}
\end{displaymath}
which we have already taken into account. This leaves the possible contractions
\begin{displaymath}
\begin{array}{l l l}
\delta^{i}_{1}=\widehat{n}^{\mu}_{i}\nabla_{\mu}\kappa &
\delta^{i}_{2}=\widehat{n}^{\nu}_{i}\widehat{\nabla}_{\mu}
\kappa^{\mu}_{i\: \nu} &
\delta^{i}_{3}=\widehat{n}^{\mu}_{i}\widehat{n}^{\nu}_{i}\widehat{n}^{\rho}_{i}
\nabla_{\mu}\kappa_{\nu \rho}
\end{array}
\end{displaymath}
plus $\delta^{j}_{1}$ etc.
The Codazzi equation 
\begin{equation}
\widehat{\nabla}_{\nu}\kappa_{i}-\widehat{\nabla}_{\mu}\kappa^{\mu}_{i\: \nu}
=R_{\mu \rho}h^{\rho}_{i\: \nu}n^{\mu}_{i}
\end{equation}
gives a link between the first two:
\begin{displaymath}
\delta^{i}_{2}=\delta^{i}_{1}-\rho^{i}_{4}
\end{displaymath}
and we discard $\delta^{i}_{2}$.

Since ${\cal I}_{ij}$ is closed and smooth, 
any term which can be written as the
divergence of a vector field $\widetilde{\nabla}_{\mu}V^{\mu}$ 
on ${\cal I}_{ij}$ can be
ignored as its integral over ${\cal I}_{ij}$ vanishes. The only vector field on
${\cal I}_{ij}$ with the appropriate scaling we can construct is
\begin{displaymath}
V^{\mu}_{i}=\gamma^{\mu \nu}\kappa^{i}_{\nu \rho}\widehat{n}^{\rho}_{i}
\end{displaymath}
Then 
\begin{displaymath}
\widetilde{\nabla}_{\mu}V^{\mu}_{i}=
\widehat{\kappa}_{i}\kappa^{i}_{\mu \nu}\widehat{n}^{\mu}_{i}
\widehat{n}^{\nu}_{i}
-\kappa^{i}_{\mu \nu}\widehat{\kappa}^{\mu \nu}_{i}
+\delta^{i}_{1}-\delta^{i}_{3}         
\end{displaymath}
The first two expressions can be written in terms of the $k$'s, so we can
discard $\delta^{i}_{3}$.

This concludes our enumeration of the terms involved in the codimension-2
contribution to $C^{(d)}_{2}$. We have finally
\begin{eqnarray}
\label{s}
{\cal S}_{ij}&=&a_{1}R +a_{2}\left( \widehat{R}_{i}+\widehat{R}_{j} \right)
+a_{3}\widetilde{R} +a_{4}R_{\mu \nu}\left( n^{\mu}_{i}\widehat{n}^{\nu}_{i}+
n^{\mu}_{j}\widehat{n}^{\nu}_{j}\right) \nonumber \\
& &+b_{1}\left( \kappa^{2}_{i}+\kappa^{2}_{j}\right)
+b_{2}\left( \hbox{tr}\kappa^{2}_{i}+\hbox{tr}\kappa^{2}_{j}\right)
+b_{3}\left(\widehat{\kappa}^{2}_{i}+\widehat{\kappa}^{2}_{j}\right) 
\nonumber \\
& &+b_{4}\left( \hbox{tr}\widehat{\kappa}^{2}_{i}
+\hbox{tr}\widehat{\kappa}^{2}_{j}\right)
+b_{5}\left( \kappa_{i}\widehat{\kappa}_{j}+\kappa_{j}\widehat{\kappa}_{i}
\right)
+c_{1}\left( \kappa_{i}\widehat{\kappa}_{i}+\kappa_{j}
\widehat{\kappa}_{j}\right)
\nonumber \\
& &+c_{2}\kappa_{i}\kappa_{j}
+c_{3}\widehat{\kappa}_{i}\widehat{\kappa}_{j}
+c_{4}\hbox{tr}\widehat{\kappa}_{i}\widehat{\kappa}_{j}
+d_{1}\left( \nabla_{\widehat{n}_{i}}\kappa_{i}
+\nabla_{\widehat{n}_{j}}\kappa_{j} \right)
\end{eqnarray}
The $a$, $b$, $c$ and $d$ coefficients are functions of $\theta$, but 
not of the dimension. The $a$'s multiply intrinsic terms
while the $b$'s and $c$'s, as we shall see, form two naturally separate groups 
when $\theta=\pi/2$.
The problem is now to evaluate these 14 coefficients by obtaining 14
independent constraints.

\section{Conformal invariance in 4 dimensions}

A great deal of information about the coefficients can be derived by
using conformal invariance. Under a conformal transformation $g_{\mu \nu}
\rightarrow 
e^{2\omega (x)}g_{\mu \nu}$, we have
\begin{eqnarray}
& &R_{\mu \nu}\rightarrow R_{\mu \nu}-g_{\mu \nu}\nabla^{2}\omega
+(d-2)\left( \nabla_{\mu}\omega \nabla_{\nu}\omega 
-g_{\mu \nu}\nabla_{\rho}\omega \nabla^{\rho}\omega
-\nabla_{\mu}\nabla_{\nu}\omega \right) \\
& &\widehat{R}^{i}_{\mu \nu}\rightarrow    
\widehat{R}^{i}_{\mu \nu}-h^{i}_{\mu \nu}\widehat{\nabla}^{2}_{i}\omega
+(d-3)\left( \widehat{\nabla}_{\mu}\omega \widehat{\nabla}_{\nu}\omega 
-h^{i}_{\mu \nu}\widehat{\nabla}_{\rho}\omega \widehat{\nabla}^{\rho}\omega
-\widehat{\nabla}_{\mu}\widehat{\nabla}_{\nu}\omega \right) \\
& &\widetilde{R}_{\mu \nu}\rightarrow    
\widetilde{R}_{\mu \nu}-\gamma_{\mu \nu}\widetilde{\nabla}^{2}\omega
+(d-4)\left( \widetilde{\nabla}_{\mu}\omega \widetilde{\nabla}_{\nu}\omega 
-\gamma_{\mu \nu}\widetilde{\nabla}_{\rho}\omega \widetilde{\nabla}^{\rho}
\omega-\widetilde{\nabla}_{\mu}\widetilde{\nabla}_{\nu}\omega \right) \\
& &\kappa^{i}_{\mu \nu}\rightarrow e^{\omega}\left( \kappa^{i}_{\mu \nu}
-h^{i}_{\mu \nu}\nabla_{n_{i}}\omega \right) \\
& &\widehat{\kappa}^{i}_{\mu \nu}\rightarrow e^{\omega}\left( 
\widehat{\kappa}^{i}_{\mu \nu}
-\gamma_{\mu \nu}\nabla_{\widehat{n}_{i}}\omega \right) 
\end{eqnarray}

We use the above to first order in $\omega$ to
work out the change in ${\cal S}_{ij}$ under a small conformal transformation
$g_{\mu \nu}\rightarrow 
e^{2\delta \omega (x)}g_{\mu \nu}$. With general
$\xi$, there is an added complication in that $C^{(d)}_{2}$ is not in fact
conformally invariant, and we must use equation (\ref{noncon}). This involves
the codimension-2 contribution to $C^{(d)}_{1}[f]$, which is known to be
\cite{dowcap}
\begin{equation}
\label{cd1i}
C^{(d)}_{1}\Big|_{\cal I}=
\sum_{(ij)}\int_{{\cal I}_{ij}}d^{d-2}x\sqrt{\gamma}\:
\frac{1}{6}\left( \frac{\pi^{2}-\theta^{2}}{\theta}\right)  f(x)
\end{equation}
The change in
$C^{(d)}_{2}$ then turns out to have the form
\begin{eqnarray}
\label{zero}
& &\Big(\delta C^{(d)}_{2}-2C^{(d)}_{1}
[J\delta \omega ]\Big)\Big|_{\cal I}   
=O(d-4)+\sum_{(ij)}\delta \int_{{\cal I}_{ij}}d^{d-2}x\sqrt{\gamma}\:
{\cal S}_{ij} \nonumber \\
& &-\eta_{d}\sum_{(ij)}\int_{{\cal I}_{ij}}d^{d-2}x\sqrt{\gamma}\: \left(
\kappa^{\mu \nu}_{i}\widehat{n}^{i}_{\mu}\delta \omega_{\nu}
-\kappa_{i}\nabla_{\widehat{n}_{i}} \delta \omega
+\kappa^{\mu \nu}_{j}\widehat{n}^{j}_{\mu}\delta \omega_{\nu}
-\kappa_{j}\nabla_{\widehat{n}_{j}} \delta \omega \right) \nonumber \\
& &-\sum_{(ij)}\int_{{\cal I}_{ij}}d^{d-2}x\sqrt{\gamma}\:
\frac{1}{3}\left( \frac{\pi^{2}-\theta^{2}}{\theta}\right) J\delta \omega
\end{eqnarray}
where $J$ is given beneath
equation (\ref{noncon}). The extra term, which vanishes in the smooth boundary
case, is equal to the boundary divergence which
comes from the conformal variation of
the volume and boundary parts of $C^{(d)}_{2}$. We keep the coefficient 
$\eta_{d}$ general since we wish briefly to discuss spin-1/2 later. 

The right-hand side of (\ref{zero}), which we must set to zero
in 4 dimensions, is expressible
in terms of the 7 linearly independent quantities
\begin{displaymath}
\begin{array}{l l}
(\kappa_{i}\nabla_{n_{i}}+\kappa_{j}\nabla_{n_{j}})\delta \omega &
(\kappa_{i}\nabla_{n_{j}}+\kappa_{j}\nabla_{n_{i}})\delta \omega \\
\: & \: \\  
(\widehat{\kappa}_{i}\nabla_{n_{i}}+\widehat{\kappa}_{j}\nabla_{n_{j}})
\delta \omega &
(\widehat{\kappa}_{i}\nabla_{n_{j}}+\widehat{\kappa}_{j}\nabla_{n_{i}})
\delta \omega \\
\: & \: \\  
(n^{\mu}_{i}n^{\nu}_{i}+n^{\mu}_{j}n^{\nu}_{j})\delta \omega_{\mu \nu} &
n^{\mu}_{i}n^{\nu}_{j}\delta \omega_{\mu \nu} \\
\: & \: \\  
(\kappa^{i}_{\mu \nu}\widehat{n}^{\mu}_{i}
+\kappa^{j}_{\mu \nu}\widehat{n}^{\mu}_{j})\delta \omega^{\nu} & \:
\end{array}
\end{displaymath}
where indices on $\delta \omega$ denote covariant derivatives on ${\cal M}$.
Any other quantity which occurs can be written in terms of these, using
(\ref{norms}) and the relations between Laplacians on ${\cal M}$, $\partial
{\cal M}_{i}$, $\partial {\cal M}_{j}$
and ${\cal I}_{ij}$. $\widetilde{\nabla}^{2} \delta \omega$ can be ignored
since it is a total divergence on $\cal I$.
Setting the coefficients of these to zero at $d=4$ gives
us 7 constraints on the $a$, $b$, $c$ and $d$. We find
\begin{eqnarray}
\label{t1}  
& &\left( 4a_{2}+6b_{1}+2b_{2}\right) \sin \theta +2b_{5}+\left( 2c_{1}+d_{1}
\right) \cos \theta =\eta_{4}\cos \theta \\
\label{t2}  
& &2b_{5}\cos \theta +3c_{2}\sin \theta +2c_{1}+d_{1}=\eta_{4} \\
\label{t3}  
& &\left( 4b_{3}+2b_{4}\right) \cos \theta +3c_{1}\sin \theta 
+2c_{3}+c_{4}=0 \\
\label{t4}  
& &-6{\overline a_{1}}-8a_{2}+4b_{3}+2b_{4}+3b_{5}\sin \theta
+\left( 2c_{3}+c_{4}\right) \cos \theta =0 \\
\label{t5}  
& &6{\overline a_{1}}+4a_{2}\left( 1+\cos^{2}\theta \right)
+\left( 2a_{4}+3d_{1}\right) \sin \theta \cos \theta =0 \\
\label{t6}  
& &\left( 6{\overline a_{1}}+8a_{2}\right) \cos \theta +\left( 2a_{4}+3d_{1}
\right) \sin \theta =0 \\
\label{t7}  
& &3d_{1}=\eta_{4}
\end{eqnarray}
where ${\overline a_{1}}$ is a correction for non-conformal coupling:
\begin{equation}
{\overline a_{1}}=a_{1}+\frac{1}{6}\left( \xi-\frac{1}{6}\right)
\left( \frac{\pi^{2}-\theta ^{2}}{\theta}\right)
\end{equation}
Also, from conformally
varying (\ref{smooth})  we find that
\begin{equation}
\eta_{d}=\frac{d-6}{90}\Rightarrow \eta_{4}=-\frac{1}{45} 
\end{equation}

\section{The smeared coefficient and conformal invariance in 6 dimensions}

There is a further conformal invariance we can use, in that \cite{bg}
\begin{equation}
\delta C^{(2k+2)}_{k}[f]=0\hbox{ as }
g_{\mu \nu}\rightarrow e^{2\delta\omega}g_{\mu \nu},
\:\:\:f\rightarrow e^{-2\delta\omega}f
\end{equation}
as long as $\Delta$ is conformally covariant
in $2k+2$ dimensions. 
 
To implement this, it is necessary to
calculate the smeared coefficient $C^{(d)}_{2}[f]$ using
(\ref{noncon}). We do not give it here since it is rather involved. Applying 
the conformal variation above in 6 dimensions with
$\xi=1/5$ then yields the following 
independent terms in the integrand of
the codimension-2 part of $\delta C^{(6)}_{2}[f]$:
\begin{displaymath}
\begin{array}{l l}
f(\kappa_{i}\nabla_{n_{i}}+\kappa_{j}\nabla_{n_{j}})\delta\omega &
f(\kappa_{i}\nabla_{n_{j}}+\kappa_{j}\nabla_{n_{i}})\delta\omega \\
\: & \: \\  
f(\widehat{\kappa}_{i}\nabla_{n_{i}}+
\widehat{\kappa}_{j}\nabla_{n_{j}})\delta\omega &
f(\widehat{\kappa}_{i}\nabla_{n_{j}}+
\widehat{\kappa}_{j}\nabla_{n_{i}})\delta\omega \\
\: & \: \\  
f(n^{\mu}_{i}n^{\nu}_{i}+n^{\mu}_{j}n^{\nu}_{j})\delta\omega_{\mu\nu}&
fn^{\mu}_{i}n^{\nu}_{j}\delta \omega_{\mu \nu} \\
\: & \: \\  
f(\kappa^{i}_{\mu \nu}\widehat{n}^{\mu}_{i}+
\kappa^{j}_{\mu \nu}\widehat{n}^{\mu}_{j})\delta\omega^{\nu} &
f\widetilde{\nabla}^{2}\delta \omega \\
\: & \: \\
(\nabla_{n_{i}}f\nabla_{n_{i}}+\nabla_{n_{j}}f\nabla_{n_{j}})\delta\omega &
(\nabla_{n_{i}}f\nabla_{n_{j}}+\nabla_{n_{j}}f\nabla_{n_{i}})\delta\omega \\
\end{array}
\end{displaymath}
All other terms can be written in terms of these, up to divergences on 
${\cal I}_{ij}$. Somewhat surprisingly, it turns out that the volume and 
boundary parts of $\delta C^{(6)}_{2}[f]$ contain no divergences, so that the
codimension-2 part vanishes by itself. We therefore set the coefficients of all
the above terms to zero. The first seven simply give us the equations we have 
already derived from 4-dimensional conformal invariance. From the remaining 
three, we get, respectively
\begin{eqnarray}
\label{t8}
& &6{\overline a_{1}}+8a_{2}+3a_{3}=0 \\
\label{t9}
& &-12 {\overline a_{1}}-16a_{2}-5(b_{1}+b_{2})\sin^{2}\theta 
+8b_{3}\left(1+\cos^{2}\theta\right) \nonumber \\
& &\:\:\:+4b_{5}\sin\theta +8c_{3}\cos \theta
+\left( 4c_{1}-\frac{11}{2}d_{1}-\frac{5}{54}\right)\cos \theta \sin \theta
=0 \\
\label{t10}
& &(-12{\overline a_{1}}-16a_{2}+16b_{3}+9b_{5}\sin \theta )\cos \theta 
\nonumber \\
& &\:\:\: +4c_{3}+\left(9c_{1}+5c_{2}\sin \theta-\frac{1}{27}
-3d_{1}\right) \sin\theta =0
\end{eqnarray}

\section {The lune}

Having extracted everything we can from conformal invariances, we now turn to
specific manifolds. Evaluation of the zeta function on lunes will turn out to
give us some more information. Lunes are particularly helpful since the angle
between adjacent boundary parts is arbitrary.
To define a $d$-lune, we start with a $2$-lune, which is the region $0\leq \phi
\leq \theta$ on a $2$-sphere of unit radius, $\phi$ being the azimuthal angle.
Higher-dimensional lunes are then defined inductively by 
$ds^{2}_{d-\hbox {\tiny lune}}
=d\chi^{2}+\sin^{2}\chi ds^{2}_{(d-1)-\hbox {\tiny lune}}$,
 $0\leq\chi\leq\pi$.

In our calculations we consider only the case $\theta =\pi /q$, where $q$ is an
integer, and analytically continue our results to all $\theta$. The
calculational procedure can be extended to $\theta =p\pi /q$, with a great deal
of added complexity. We are assuming that the spectral coefficients are smooth
functions of $\theta$ -- this is the case for $C^{(d)}_{1}$ (\ref{cd1i}).
Some work has been done by Cook \cite{tony} on the case where $\theta
/\pi$ is irrational. The results agree with analytic continuation.

We will consider $2\leq d \leq 5$. Since ${\cal S}_{ij}$ (\ref{s}) is quadratic
in $d$ on a spherical domain, we can derive 
at most three independent constraints, and stopping
at $d=4$ turns out to be sufficient. $d=5$ provides a check of the equations.
The procedure is to use the eigenvalues and degeneracies
to calculate the zeta function and derive the heat-kernel 
coefficients via (\ref{czeta}). The zeta function on orbifolded spheres, of 
which the lune is an example, has been calculated \cite{candd}, and we do not 
go into detail here.
For a general coupling $\xi$, we find
\begin{eqnarray}
C^{(2)}_{2}&=&\frac{\pi}{720q}\left( 8q^{4}+20q^{2}-7\right)
+\frac{\pi \alpha^{2}}{6q}\left( 2q^{2}-1\right) +\frac{\pi \alpha^{4}}{q} \\
C^{(3)}_{2}&=&\frac{\pi^{2} \alpha^{2}}{3q}\left( q^{2}-1\right)
+\frac{\pi^{2} \alpha^{4}}{2q} \\
C^{(4)}_{2}&=&\frac{\pi^{2}}{360q}\left( 51-60q^{2}-8q^{4}\right)
+\frac{\pi^{2}\alpha^{2}}{3q}\left( 2q^2-3\right)
+\frac{2\pi^{2}\alpha^{4}}{3q} \\
C^{(5)}_{2}&=&\frac{\pi^{3}}{45q}\left( 11-10q^{2}-q^{4}\right)
+\frac{\pi^{3}\alpha^{2}}{3q}\left( q^{2}-2\right) 
+\frac{\pi^{3}\alpha^{4}}{4q} 
\end{eqnarray}
where
\begin{displaymath}
\alpha^{2}=d(d-1)\left[\frac{d-1}{4d}-\xi\right]
\end{displaymath}

In using this data to obtain information about ${\cal S}_{ij}$, we need to
separate off the codimension-2 contribution to $C^{(d)}_{2}$. On the lune all
extrinsic curvatures vanish, as does $\nabla_{n}R$, 
so the boundary contribution is zero. Since the
volume part is proportional to $1/q$, we have
\begin{displaymath}
C^{(d)}_{2}(q)\Big|_{\cal I} =C^{(d)}_{2}(q)-\frac{1}{q}C^{(d)}_{2}(q=1)
\end{displaymath}
The boundary intersection ${\cal I}$ 
on a lune is a $(d-2)\:$-sphere of unit radius
(2 points, each of
content 1, in the case $d=2$). The boundaries are $(d-1)$-hemispheres of unit
radius.
The only nonvanishing geometrical quantities are
\begin{displaymath}
\begin{array}{l l l}
R=d(d-1) & \widehat{R}_{i}=\widehat{R}_{j}=(d-1)(d-2) & 
\widetilde{R}=(d-2)(d-3) \\
\end{array}
\end{displaymath}
$R_{\mu \nu}n^{\mu}_{i}\widehat{n}^{\nu}_{i}$ and
$R_{\mu \nu}n^{\mu}_{j}\widehat{n}^{\nu}_{j}$ vanish since the lune is an 
Einstein
space $R_{\mu \nu}=(d-1)g_{\mu \nu}$ and the two normals are orthogonal.

Equating equation (\ref{s}) with our expressions for 
$C^{(d)}_{2}\Big|_{\cal I}$ on the $d$-lune, we obtain the constraints
\begin{eqnarray*}
a_{1}&=&\frac{\pi}{36}(6\xi -1)\left( 
\frac{\theta}{\pi}-\frac{\pi}{\theta} \right)+  
\frac{\pi}{360}\left( \frac{\pi^{3}}{\theta^{3}}
-\frac{\theta}{\pi}\right) \\
3a_{1}+2a_{2}&=&\frac{\pi}{12}(6\xi -1)\left( 
\frac{\theta}{\pi}-\frac{\pi}{\theta} \right)  \\
6a_{1}+6a_{2}+a_{3}&=&\frac{\pi}{6}(6\xi -1)\left( 
\frac{\theta}{\pi}-\frac{\pi}{\theta} \right)+  
\frac{\pi}{360}\left( \frac{\theta}{\pi}-\frac{\pi^{3}}{\theta^{3}}\right)
\end{eqnarray*}
from 2, 3 and 4 dimensions respectively. In each case we can conveniently
replace $a_{1}$ by ${\overline a}_{1}$ and remove the $(6\xi -1)$ term.
We now have $a_{1}$, $a_{2}$ and $a_{3}$ for all $\theta$:
\begin{eqnarray}
\label{a1}
& &{\overline a}_{1}=\frac{\pi}{360}\left( \frac{\pi^{3}}{\theta^{3}}-
\frac{\theta}{\pi}\right) \\
\label{a2}        
& &a_{2}=-\frac{\pi}{240}\left( \frac{\pi^{3}}{\theta^{3}}-
\frac{\theta}{\pi}\right) \\
\label{a3} 
& &a_{3}=\frac{\pi}{180}\left( \frac{\pi^{3}}{\theta^{3}}-
\frac{\theta}{\pi}\right) 
\end{eqnarray}
>From 5 dimensions we get
\begin{displaymath}
10{\overline a}_{1}+12a_{2}+3a_{3}=
-\frac{\pi}{180}\left( \frac{\pi^{3}}{\theta^{3}}-
\frac{\theta}{\pi}\right) 
\end{displaymath}
as a check of (\ref{a1}), (\ref{a2}) and (\ref{a3}). 

The conformal variation equations (\ref{t5}) and (\ref{t6}) give 
$3{\overline a}_{1}+2a_{2}=0$, in agreement with the above. In addition, the 
lune results agree with equation (\ref{t8}).
This is
encouraging, although we have less information than we hoped for, i.e. lunes
have only given us 1 new constraint.
               
We now have all the $a$ coefficients for general $\theta$, since
from (\ref{t6}), (\ref{t7})
\begin{equation}
\label{a4}
a_{4}=\frac{1}{90}+\frac{\pi}{120}\left(\frac{\pi^{3}}{\theta^{3}}
-\frac{\theta}{\pi}\right)\cot\theta
\end{equation}

\section{A further constraint}

We can in fact make the
situation a bit simpler by noticing that $k^{i}_{2}$ is conformally invariant
for {\it all} $d$, so we can consistently set this quantity to zero. This
restriction makes sense since 
$k^{i}_{2}$ vanishes on most spaces we are liable to deal
with, e.g. the hemiball, lune and cylinder. $k^{i}_{2}$ measures the rate at
which $n_{i}$ rotates in the direction of $\widehat{n}_{i}$ as we move around 
on
${\cal I}_{ij}$ -- an example of a boundary where $k^{i}_{2}$ does not vanish 
is a rectangle twisted so that each edge forms a helix.

If we set $k^{i}_{2}=k^{j}_{2}=0$, 
there are now only 13 degrees of freedom, and we can
remove any one of the terms with $b$ or $c$ coefficients from (\ref{s}). We
choose to get rid of $\kappa_{i}\widehat{\kappa}_{j}$ and 
$\kappa_{j}\widehat{\kappa}_{i}$, so we set $b_{5}$ to zero.

In the case where this quantity did not vanish, due to its conformal invariance
in all dimensions we would not be able to find the extra degree of freedom in 
the coefficients except by calculating the zeta function on a manifold where 
the measure of ``twist'' we have described is non-zero. As far as we can see,
this would present a very difficult task and for the moment it is necessary to
make this constraint if we wish to complete the calculation.

\section{Right-angled edges and product manifolds}

We can go further if we limit ourselves to the case $\theta=\pi/2$. Again, many
of the manifolds we come across have this property. We 
also maintain the constraint
we have made above, that $k^{i}_{2}=0$. For $\theta=\pi/2$, this 
condition takes the simple form
\begin{equation}
\label{notwist}
\left( \kappa_{i}-\widehat{\kappa}_{j}\right)^{2} =
\hbox{tr}\kappa^{2}_{i}-\hbox{tr}\widehat{\kappa}^{2}_{j} 
\end{equation}
Our 11 equations, now for only 13 coefficients, become
\begin{eqnarray}
\label{info1}
& &{\overline a_{1}}=\frac{\pi}{48}\:\:\:\:\:\:\:\:\:\:\:\:\:
a_{2}=-\frac{\pi}{32}\:\:\:\:\:\:\:\:\:\:\:\:\:
a_{3}=\frac{\pi}{24}\:\:\:\:\:\:\:\:\:\:\:\:\:
a_{4}=\frac{1}{90}\nonumber \\
& &3b_{1}+b_{2}=\frac{\pi}{16}\:\:\:\:\:\:
2b_{3}+b_{4}=-\frac{\pi}{16}\:\:\:\:\:\:
b_{1}+5b_{2}-8b_{3}=\frac{\pi}{4}\nonumber \\
& &2c_{1}+3c_{2}=-\frac{2}{135}\:\:\:\:\:\:\:
3c_{1}+2c_{3}+c_{4}=0\:\:\:\:\:\:\:
d_{1}=-\frac{1}{135}\nonumber \\
& &9c_{1}+5c_{2}+4c_{3}=\frac{2}{135}
\end{eqnarray}

We can derive the last two constraints we need by considering a product 
manifold ${\cal M}={\cal M}_{1}\times{\cal M}_{2}$, with an operator 
$\Delta =-\left(\nabla^{2}_{1}+\nabla^{2}_{2}\right)
+\xi(R_{1}+R_{2})$. ${\cal M}_{1}$ and ${\cal M}_{2}$ have smooth boundaries.
${\cal M}$ has a codimension-2 submanifold $\partial {\cal M}_{1}
\times\partial{\cal M}_{2}$ with $\theta =\pi/2$, and obeys (\ref{notwist}).

>From (\ref{hk}), it is easily shown that, for scalar fields with Dirichlet 
conditions
\begin{equation}
\label{prod}
C^{\cal M}_{2}=C^{{\cal M}_{1}}_{0}C^{{\cal M}_{2}}_{2}
+C^{{\cal M}_{1}}_{1/2} C^{{\cal M}_{2}}_{3/2}
+C^{{\cal M}_{1}}_{1} C^{{\cal M}_{2}}_{1}
+C^{{\cal M}_{1}}_{3/2} C^{{\cal M}_{2}}_{1/2}
+C^{{\cal M}_{1}}_{2} C^{{\cal M}_{2}}_{0}     
\end{equation}
The codimension-2 part of the left hand side is, in terms of geometrical 
quantities on ${\cal M}_{1}$ and ${\cal M}_{2}$
\begin{eqnarray*}
\int_{\partial {\cal M}_{1}\times \partial{\cal M}_{2}}
d^{d-2}x\sqrt{h_{1}h_{2}}\:\Big[
(a_{1}+a_{2})(R_{1}+R_{2}) 
+(a_{2}+a_{3})\left( \widehat{R}_{1}+\widehat{R}_{2}\right)\\ 
+(b_{1}+b_{3})\left(\kappa^{2}_{1}+\kappa^{2}_{2}\right) 
+(b_{2}+b_{4})\left(\hbox{tr}\kappa^{2}_{1}
+\hbox{tr}\kappa^{2}_{2}\right) 
+(2c_{1}+c_{2}+c_{3})\kappa_{1}\kappa_{2}\Big]
\end{eqnarray*}

The expressions for 
$C^{(d)}_{k}$ for $k=0$, $1/2$, $1$, $3/2$ are well-known
\cite{minak,kennedy}:
\begin{eqnarray}
& &C^{(d)}_{0}=\int_{\cal M}d^{d}x\sqrt{g}\\
& &C^{(d)}_{1/2}=-\frac{\sqrt{\pi}}{2}
\int_{\partial {\cal M}}d^{d-1}x\sqrt{h}\\
& &C^{(d)}_{1}=\left(\frac{1}{6}-\xi\right)
\int_{\cal M}d^{d}x\sqrt{g}\:R
+\frac{1}{3}\int_{\partial {\cal M}}d^{d-1}x\sqrt{h}\kappa\\
& &C^{(d)}_{3/2}=\frac{\sqrt{\pi}}{192}\int_{\partial {\cal M}}d^{d-1}x
\sqrt{h}\:\left[ 6\hbox{tr}\kappa^{2}-3\kappa^{2}
+12(8\xi-1)R-4\widehat{R}\right]
\end{eqnarray}
The codimension-2 part of the right-hand side of (\ref{prod}) then comes from 
the middle three terms, and is 
\begin{eqnarray*}
\int_{\partial {\cal M}_{1}\times \partial{\cal M}_{2}}
d^{d-2}x\sqrt{h_{1}h_{2}}\:\Bigg[
\frac{1}{9}\kappa_{1}\kappa_{2}-\frac{\pi}{384}\bigg[
6\left(\hbox{tr}\kappa^{2}_{1}+\hbox{tr}\kappa^{2}_{2}\right)\\
-3\left(\kappa^{2}_{1}+\kappa^{2}_{2}\right)
+12(8\xi-1)(R_{1}+R_{2})-4\left(\widehat{R}_{1}+\widehat{R}_{2}\right)\bigg]
\Bigg]
\end{eqnarray*}
Equating the two sides, we find
\begin{eqnarray}
{\overline a}_{1}+a_{2}=-\frac{\pi}{96}& &
a_{2}+a_{3}=\frac{\pi}{96}\nonumber \\
b_{1}+b_{3}=\frac{\pi}{128}& &
b_{2}+b_{4}=-\frac{\pi}{64}\nonumber \\
2c_{1}+c_{2}+c_{3}=\frac{1}{9}
\end{eqnarray}
The first two of these agree with (\ref{info1}). The second two give one new 
constraint, completing the information for the $b$'s, and one in agreement with
the rest. The last completes the information for the $c$'s. Our final result is
\begin{eqnarray}
\label{result}
C^{(d)}_{2}\Big|_{\cal I}&=&
\sum_{(ij)}\int_{{\cal I}_{ij}}d^{d-2}x\sqrt{\gamma}\:
\Bigg\{\frac{\pi}{384}\bigg[
24(1-4\xi)R-12\left(\widehat{R}_{i}+\widehat{R}_{j}\right)+16\widetilde{R}
\nonumber\\
& &+24\left(\hbox{tr}\kappa^{2}_{i}+\hbox{tr}\kappa^{2}_{j}\right)
+3\left(\widehat{\kappa}^{2}_{i}+\widehat{\kappa}^{2}_{j}\right)
-30\left(\hbox{tr}\widehat{\kappa}^{2}_{i}+\hbox{tr}\widehat{\kappa}^{2}_{j}
\right)\bigg]\nonumber\\
& &+\frac{1}{270}\bigg[
3R^{\mu\nu}\left(n^{\mu}_{i}\widehat{n}^{\nu}_{i}
+n^{\mu}_{j}\widehat{n}^{\nu}_{j}\right)
-344\left(\kappa_{i}\widehat{\kappa}_{i}
+\kappa_{j}\widehat{\kappa}_{j}\right)
+228\kappa_{i}\kappa_{j}\nonumber\\
& &+490\widehat{\kappa}_{i}\widehat{\kappa}_{j}
+52\hbox{tr}\widehat{\kappa}_{i}\widehat{\kappa}_{j}
-2\left(\nabla_{\widehat{n}_{i}}\kappa_{i}
+\nabla_{\widehat{n}_{j}}\kappa_{j}\right)
\bigg]\Bigg\}
\end{eqnarray}
We note that this, together with equation
(\ref{smooth}), can be extended to the non-zero mass case, 
where the operator becomes
\begin{displaymath}
\Delta=-\nabla^{2}+\xi R-m^{2}
\end{displaymath}
simply by replacing $\xi R$ by $\xi R-m^{2}$.

\section{The smeared coefficient and cocycle function}
For completeness, we give the smeared coefficient with $\theta=\pi/2$, where 
(\ref{notwist}) applies. The expression for the smooth boundary case is 
well-known. Using (\ref{smooth}), (\ref{result}) and (\ref{noncon}), we find
\begin{eqnarray}
\label{smear}
C^{(d)}_{2}[f]&=&\int_{\cal M}d^{d}x\sqrt{g}\:
\bigg[ {\cal U}f +\frac{1}{6}\left(\frac{1}{5}-\xi\right)
R\nabla^{2}f\bigg]\nonumber\\
& &+\sum_{i}\int_{\partial {\cal M}_{i}}d^{d-1}x\sqrt{h}\:
\bigg[ {\cal T}f+\frac{1}{3}\left(\xi-\frac{3}{20}\right)R\nabla_{n}f
+\frac{1}{15}\kappa\nabla^{2}f\nonumber\\
& &-\frac{1}{210}\kappa^{2}\nabla_{n}f
+\frac{1}{42}\hbox{tr}\kappa^{2}\nabla_{n}f
-\frac{1}{12}\nabla_{n}\nabla^{2}f\bigg]\nonumber\\
& &+\sum_{(ij)}\int_{{\cal I}_{ij}}d^{d-2}x\sqrt{\gamma}\:
\Bigg\{{\cal S}_{ij}f+\frac{\pi}{64}\bigg[
4\left(\kappa_{i}\nabla_{n_{i}}+\kappa_{j}\nabla_{n_{j}}\right)f\nonumber\\
& &-9\left(\widehat{\kappa}_{i}\nabla_{n_{j}}
+\widehat{\kappa}_{j}\nabla_{n_{i}}\right)f
+4\left(n^{\mu}_{i}n^{\nu}_{i}+n^{\mu}_{j}n^{\nu}_{j}\right)f_{\mu\nu}
\bigg]\nonumber\\
& &+\frac{1}{270}\bigg[119\left(\kappa_{i}\nabla_{n_{j}}
+\kappa_{j}\nabla_{n_{i}}\right)f
-146\left(\widehat{\kappa}_{i}\nabla_{n_{i}}+\widehat{\kappa}_{j}\nabla_{n_{j}}
\right)f\nonumber\\
& &-2n^{\mu}_{i}n^{\nu}_{j}f_{\mu\nu}
-5\left(\kappa^{i}_{\mu\nu}\widehat{n}^{\mu}_{i}
+\kappa^{j}_{\mu\nu}\widehat{n}^{\mu}_{j}\right)f^{\nu}\bigg]\Bigg\}
\end{eqnarray}
where ${\cal U}$ and ${\cal T}$ are the volume and boundary parts of 
the standard smooth expression for $C^{(d)}_{2}[1]$, (\ref{smooth}).

We can use this to calculate the change in the effective action,
or cocycle function, 
in 4 dimensions for a conformal field theory ($\xi=1/6$).
It can be shown using zeta function 
techniques and $W=-\frac{1}{2}\zeta'(0)$
that under a small conformal variation, 
\begin{displaymath}
\delta W=-\frac{1}{(4\pi)^{d/2}}C^{(d)}_{d/2}[\delta\omega]
\end{displaymath}
The difference in effective action for two metrics $g$ and 
${\overline g}=e^{2\omega}g$ is then given by
\begin{equation}
W\left[{\overline g}\right]-W\left[g\right]=
-\frac{1}{(4\pi)^{d/2}}\int_{u=0}^{1}C^{(d)}_{d/2}\left[e^{2u\omega}g;
\omega du\right]
\end{equation}
Again the volume and boundary parts are known
\cite{schofield}, the result being
\begin{eqnarray}
\label{cosmooth}
\Big(W\left[{\overline g}\right]-W[g]\Big)\Big|_{V,B}
&=&\frac{1}{2880\pi^{2}}\int_{\cal M}d^{4}x\sqrt{g}\:
\left[-180{\cal U}\omega-\omega\nabla^{2}R \right. \nonumber \\
& &\left.-2R^{\mu \nu}\omega_{\mu}\omega_{\nu}
+4\omega^{\mu}\omega_{\mu}\nabla^{2}\omega
+2(\omega^{\mu}\omega_{\mu})^{2}
+3(\nabla^{2}\omega)^{2}\right] \nonumber \\        
& &+\frac{1}{5760\pi^{2}}\sum_{i} 
\int_{\partial {\cal M}_{i}} d^{3}x\sqrt{h}\:
\bigg[-360 {\cal T}\omega-2\omega\nabla_{n}R\nonumber \\
& &+\nabla_{n}\omega \left( \frac{12}{7}\kappa^{2}
-\frac{60}{7}\hbox{tr}\kappa^{2}
+12\nabla^{2}\omega
+8\omega^{\mu}\omega_{\mu}\right) \nonumber \\
& &-\frac{4}{7}\kappa (\nabla_{n}\omega)^{2}
-\frac{16}{21}(\nabla_{n}\omega)^{3}
-24\kappa \nabla^{2}\omega
-4\kappa^{\mu \nu}\omega_{\mu}\omega_{\nu} \nonumber \\
& &-20\kappa \omega^{\mu}\omega_{\mu}
+30\nabla_{n}(\nabla^{2}\omega +\omega^{\mu}\omega_{\mu}) \bigg]
\end{eqnarray}

The calculation of the edge part is fairly easy, and we find
\begin{eqnarray}
\label{coedge}
\Big(W\left[{\overline g}\right]-W[g]\Big)\Big|_{\cal I}&=&
-\frac{1}{16\pi^{2}}
\sum_{(ij)}\int_{\cal I}d^{2}x\sqrt{\gamma}\:
\Bigg\{{\cal S}_{ij}\omega+\frac{\pi}{64}\bigg[
4\left(\kappa_{i}\nabla_{n_{i}}+\kappa_{j}\nabla_{n_{j}}\right)
\omega\nonumber\\
& &-9\left(\widehat{\kappa}_{i}\nabla_{n_{j}}
+\widehat{\kappa}_{j}\nabla_{n_{i}}\right)\omega
+4\left(n^{\mu}_{i}n^{\nu}_{i}+n^{\mu}_{j}n^{\nu}_{j}\right)
\omega_{\mu\nu}\nonumber\\
& &+3\left[(\nabla_{n_{i}}\omega)^{2}
+(\nabla_{n_{j}}\omega)^{2}\right]
-\frac{8}{3}\omega\widetilde{\nabla}^{2}\omega\bigg]\nonumber\\
& &+\frac{1}{270}\bigg[119\left(\kappa_{i}\nabla_{n_{j}}
+\kappa_{j}\nabla_{n_{i}}\right)\omega
-146\left(\widehat{\kappa}_{i}\nabla_{n_{i}}+\widehat{\kappa}_{j}\nabla_{n_{j}}
\right)\omega\nonumber\\
& &-2n^{\mu}_{i}n^{\nu}_{j}\omega_{\mu\nu}
-5\left(\kappa^{i}_{\mu\nu}\widehat{n}^{\mu}_{i}
+\kappa^{j}_{\mu\nu}\widehat{n}^{\mu}_{j}\right)\omega^{\nu}
-58\nabla_{n_{i}}\omega\nabla_{n_{j}}\omega\bigg]\Bigg\}
\end{eqnarray} 
 
This formula can be used to find the effective action on a manifold if the 
value on a manifold to which it is conformally related is known. Previously, in
4 dimensions, we could only consider spaces
with smooth boundaries -- we can now do calculations where an edge is present.

For example, a hemicap of a 4-sphere of unit radius
(${\overline g}$)
is related to the hemiball ($g$) by the 
conformal transformation
\begin{displaymath}
{\overline g}_{\mu\nu}=e^{2\omega}g_{\mu\nu},\:\:\:
\omega=\ln\left(\frac{2}{1+r^{2}}\right)
\end{displaymath}
r being the radial coordinate on the hemiball, 
where the radius $a$ of the hemicap and the colatitude $\Theta$ of the hemicap 
boundary are related by
\begin{displaymath}
a=\tan\frac{\Theta}{2}
\end{displaymath}
If $a=1$, then the hemicap is a quarter-sphere. We find
\begin{eqnarray}
W_{\hbox{\tiny hemiball}}(a)-W_{\hbox{\tiny 1/4-sphere}}&=&
\frac{19}{1440}\ln a-\frac{1}{360}\ln 2
-\frac{6751}{241920}\nonumber \\
W_{\hbox{\tiny hemicap}}(\Theta)-W_{\hbox{\tiny 1/4-sphere}}&=&
\frac{1}{96}\ln\left(1+\cos\Theta\right)
+\frac{1}{1440}\bigg[19\ln\tan\Theta/2\nonumber \\
& &+\frac{5}{8}\cos\Theta\left(139+63\cos\Theta\right)
-\frac{1399}{168}\cos^{3}\Theta\bigg]
\end{eqnarray}
Expression for the quartersphere effective action have been derived in
\cite{ffd,dowsph2}.

\section{Spin-1/2}

A similar calculation should be possible if $\Delta$ is the squared Dirac 
operator
\begin{displaymath}
\Delta=\left(i\gamma^{\mu}\nabla_{\mu}\right)^{2}
=-g^{\mu\nu}\nabla_{\mu}\nabla_{\nu}+\frac{1}{4}R
\end{displaymath}
as long as the boundary conditions are local. A suitable set are
the mixed conditions \cite{luck1,luck2}
\begin{displaymath}
P_{-}\psi\Big|_{\partial{\cal M}}=0,\:\:\:
\left(\nabla_{n}-\frac{1}{2}\kappa\right)\psi\Big|_{\partial{\cal M}}=0
\end{displaymath}
for the projection operators $P_{\pm}=\frac{1}{2}(1\pm 
i\gamma^{5}\gamma^{\mu}n_{\mu})$.

In this case the volume and boundary parts of $C_{2}$ are known 
\cite{moss1,moss2,thesis}. 
${\cal S}_{ij}$ has the form (\ref{s}), and the equations we derive from 
conformal invariance in 4 dimensions are identical to (\ref{t1}) to 
(\ref{t7}), with the exception that
\begin{displaymath}
\eta_{d}=\frac{2d+3}{180}{\cal D}
\end{displaymath}
where ${\cal D}$ is the dimension of the spinor space, and ${\overline a}_{1}
=a_{1}$ since there is no variable coupling to take into account.

>From direct calculation on lunes we find
\begin{eqnarray}
C^{(2)}_{2}&=&-\frac{\pi{\cal D}}{720q}\left( 7q^{4}+10q^{2}+7\right) \\
C^{(3)}_{2}&=&0 \\
C^{(4)}_{2}&=&\frac{\pi^{2}{\cal D}}{360q}\left( 7q^{4}+30q^{2}+51\right) \\
C^{(5)}_{2}&=&\frac{\pi^{3}{\cal D}}{360q}\left( 7q^{4}+40q^{2}+88\right)
\end{eqnarray}       
so that (from the first 3 of these)
\begin{eqnarray}
a_{1}&=&-\frac{\pi{\cal D}}{2880}\left( 7\frac{\pi^{3}}{\theta^{3}}
+10\frac{\pi}{\theta} -17\frac{\theta}{\pi}\right) \\
a_{2}&=&\frac{\pi{\cal D}}{1920}\left( 
7\frac{\pi^{3}}{\theta^{3}} +10\frac{\pi}{\theta}
-17\frac{\theta}{\pi}\right) \\
a_{3}&=&\frac{7\pi{\cal D}}{1440}\left( \frac{\theta}{\pi}-
\frac{\pi^{3}}{\theta^{3}}\right)
\end{eqnarray}
>From 5 dimensions we get
\begin{displaymath}
10a_{1}+12a_{2}+3a_{3}=\frac{\pi{\cal D}}{1440}\left(
7\frac{\pi^{3}}{\theta^{3}}+40\frac{\pi}{\theta}-47\frac{\theta}{\pi}\right)
\end{displaymath}
in agreement with the other relations.
As for the scalar case, we have $3{\overline a}_{1}+2a_{2}=0$, in agreement
with the conformal equations. Additionally,
\begin{equation}
a_{4}=-\frac{11{\cal D}}{360}-\frac{\pi{\cal D}}{960}
\left(7\frac{\pi^{2}}{\theta^{2}}
+10\frac{\pi}{\theta}-17\frac{\theta}{\pi}\right)\cot\theta
\end{equation}

Unfortunately, this is as far as we can go with the techniques used in this 
paper. Conformal invariance of the smeared coefficient in 6 dimensions does not
apply -- this is to do with the fact that the squared Dirac operator is not in 
fact conformally covariant; it is only a power of a conformally covariant 
operator. Additionally, on the product of two manifolds with boundaries, the 
heat-kernel does not turn out to be a simple product, i.e. equation 
(\ref{prod}) does not apply. This is a result of the boundary conditions.

\section{Robin boundary conditions}

A conformally invariant boundary condition, alternative to the Dirichlet case,
is the Robin condition
\begin{equation}
\label{rob}
\left(\nabla_{n}\phi-\psi\phi\right)\Big|_{\partial {\cal M}}=0
\end{equation}
for some boundary function $\psi$,
where under a conformal transformation we set
\begin{equation}
\psi\rightarrow e^{-\omega}\left[\psi-\frac{1}{2}(d-2)\nabla_{n}\omega\right]
\end{equation}
so that (\ref{rob}) is conformally invariant.

A general expression for the integrand of the codimension-2 contribution is
\begin{eqnarray}
\label{sr}
{\cal S}_{ij}&=&{\cal S}^{N}_{ij}
+u_{1}\left(\kappa_{i}\psi_{j}+\kappa_{j}\psi_{i}\right)
+u_{2}\left(\widehat{\kappa}_{i}\psi_{j}+\widehat{\kappa}_{j}\psi_{i}\right)
\nonumber\\
& &+u_{3}\left(\psi^{2}_{i}+\psi^{2}_{j}\right)\nonumber 
+v_{1}\left(\kappa_{i}\psi_{j}+\kappa_{j}\psi_{i}\right)
+v_{2}\left(\widehat{\kappa}_{i}\psi_{i}+\widehat{\kappa}_{j}\psi_{j}\right)
\nonumber\\
& &+v_{3}\psi_{i}\psi_{j}
+w_{1}\left(\nabla_{\widehat{n}_{i}}\psi_{i}+\nabla_{\widehat{n}_{j}}
\psi_{j}\right)
\end{eqnarray}
where the Neumann ($\psi_{i}=\psi_{j}=0$) expression 
${\cal S}^{N}_{ij}$ has the same form as the Dirichlet expression (\ref{s}),
although in general the $a$, $b$, $c$ and $d$ constants will have different 
values.

Going through the procedure we have detailed for the Dirichlet case,
this time sticking with $\theta=\pi/2$, yields 
\begin{eqnarray}
& &{\overline a}_{1}=\frac{\pi}{48}\:\:\:\:\:\:
a_{2}=-\frac{\pi}{32}\:\:\:\:\:\:
a_{3}=\frac{\pi}{24}\:\:\:\:\:\:
a_{4}=\frac{1}{90}\nonumber\\
& &b_{1}=\frac{\pi}{16}\:\:\:\:\:
b_{2}=\frac{\pi}{8}\:\:\:\:\:
b_{3}=-\frac{5\pi}{128}\:\:\:\:\:
b_{4}=-\frac{7\pi}{64}\nonumber\\
& &u_{1}=-\frac{\pi}{2}\:\:\:\:\:\:\:\:
u_{2}=\frac{\pi}{4}\:\:\:\:\:\:\:\:
u_{3}=\frac{\pi}{2}\:\:\:\:\:\:\:\:
v_{3}=4
\end{eqnarray}
with 7 equations for the remaining 8 constants:
\begin{eqnarray}
& &2c_{1}+3c_{2}+d_{1}+v_{1}=-\frac{1}{45}\\
& &3c_{1}+2c_{3}+c_{4}+v_{2}=0\\
& &3d_{1}+w_{1}=-\frac{1}{45}\\
& &3v_{1}+2v_{2}+v_{3}+w_{1}=0\\
& &9c_{1}+5c_{2}+4c_{3}-3d_{1}+\frac{9}{2}v_{1}+4v_{2}
+v_{3}-\frac{1}{6}w_{1}
=\frac{1}{27}\\
& &2c_{1}+c_{2}+c_{3}=\frac{1}{9}\\
& &v_{1}+v_{2}=-\frac{2}{3}
\end{eqnarray}
with checks.

To acquire the final piece of information, we consider the specific case of a 
square with Neumann conditions on three sides and a non-zero boundary function 
on the fourth. While this analysis would be difficult for general $\psi$, it 
is possible to calculate the eigenvalues and zeta function for a small 
perturbation $\psi=\epsilon f$, $\epsilon\ll 1$. Our results will be correct 
to first order in $\epsilon$. This is sufficient since we will be able to 
obtain the coefficient of the $\nabla_{\widehat{n}_{i}}\psi_{i}$ term in 
$C_{2}$.

It can be shown that under such a perturbation, in general the eigenvalues 
become
\begin{equation}
\lambda=\lambda_{N}+\epsilon\eta,\:\:\:
\eta=\int_{\partial{\cal M}}d^{d-1}x\sqrt{h}\:|\phi_{N}|^2\,f
\end{equation}
where the $\lambda_{N}$ and $\phi_{N}$ are the 
eigenvalues and normalized eigenfunctions for $\epsilon=0$ 
(Neumann conditions).

In the case of a square of side $\pi$, with $\psi=\epsilon f(y)$ on the $x=0$ 
boundary,
\begin{eqnarray*}
\eta_{mn}=\frac{4}{\pi^{2}}\int_{0}^{\pi}dy\:f\cos^{2}ny,& &
\eta_{0n}=\frac{1}{2}\eta_{mn}\\
\eta_{m0}=\frac{2}{\pi^{2}}\int_{0}^{\pi}dy\:f,& &
\eta_{00}=\frac{1}{2}\eta_{m0},\:\:\:
m,n>0
\end{eqnarray*}
where the $m$, $n$ label the eigenfunctions $\phi_N\sim \cos mx\cos ny$ with 
the eigenvalues $m^{2}+n^{2}$.

We will find it convenient to write
\begin{displaymath}
\eta_{mn}=A+\frac{B}{n^{2}}+O\left(\frac{1}{n^{4}}\right)
\end{displaymath}
where, from integration by parts,
\begin{displaymath}
A=\frac{2}{\pi^{2}}\int_{0}^{\pi}dy\:f,\:\:\:\:\:
B=\frac{1}{2\pi^{2}}\left[f'(\pi)-f'(0)\right]
\end{displaymath}
Then to first order in $\epsilon$,
\begin{eqnarray*}
\zeta(s)&=&\zeta_{N}(s)-\frac{1}{2}s\epsilon
\sum_{m=-\infty}^{\infty}\sum_{n=1}^{\infty}
\left[A+\frac{B}{n^{2}}+O\left(\frac{1}{n^{4}}\right)\right]
\left(m^{2}+n^{2}\right)^{-(s+1)}\\
& &-2s\epsilon\eta_{00}\zeta_{R}(2s+2)+(\eta_{00}\epsilon)^{-s}
\end{eqnarray*}

The evaluation of the sum is an old, well-known procedure 
\cite{epstein} used in calculating 
the zeta function on cylinders -- see for example 
\cite{ffd,ken2}. The standard result is
\begin{eqnarray}
\label{ep}
\sum_{m=-\infty}^{\infty}\sum_{n=1}^{\infty}
g(n)\left(m^{2}+n^{2}\right)^{-u}
=\frac{\Gamma(u-1/2)\sqrt{\pi}}{\Gamma(u)}
\sum_{n=1}^{\infty}g(n)n^{1-2u}\nonumber \\
+\frac{2\sqrt{\pi}}{\Gamma(u)}\sum_{m=1}^{\infty}\sum_{n=1}^{\infty}
\int_{0}^{\infty}dt\:t^{u-3/2}g(n)e^{-n^{2}t}e^{-m^{2}\pi^{2}/t}
\end{eqnarray}

Since we are interested in $C_{2}=-4\pi\zeta(-1)$ we wish to evaluate the 
above expression for $u=0$. If $g(n)$ is a finite sum of powers of $1/n^{2}$, 
the second term is zero, since the sum
over $m$ and $n$ converges and 
$\lim_{u\rightarrow 0}\Gamma(u)=1/u$. The first picks out the $1/n^{2}$ 
term from $g(n)$ since $\lim_{u\rightarrow 0}\zeta_{R}(1+2u)=1/2u$, and 
$\zeta_{R}$ is finite for all other arguments. We are left with
\begin{eqnarray*}
& &\zeta(-1)=\zeta_{N}(-1)-\frac{1}{2}\epsilon B\pi
+2\epsilon\eta_{00}\zeta_{R}(0)+\epsilon\eta_{00}\\
&\Rightarrow& C_{2}=C^{N}_{2}+\epsilon\left[f'(\pi)-f'(0)\right]
\end{eqnarray*}
since $\zeta_{R}(0)=-1/2$. Comparing with (\ref{sr}) immediately gives 
us the required number
\begin{equation}
w_{1}=-1
\end{equation}
We note that there may possibly be functions $f(y)$ for which this analysis 
does not apply since the sum in (\ref{ep}) diverges. However it is of course 
sufficient to consider only one special case -- certainly our arguments hold 
for {\it eg.} polynomial $f(y)$, where the expansion of $g(n)$ terminates. 

Other heat-kernel coefficients may be evaluated in this way. For example, 
setting $s=0$, it is the constant term in $g(n)$ which contributes, and
\begin{displaymath}
\zeta(0)=\zeta_{N}(0)-\frac{\pi}{4}\epsilon A+1
\end{displaymath}
In the Neumann case there is a zero mode which is not included in the zeta 
function, so that $C^{N}_{1}=4\pi(\zeta_{N}(0)+1)$, 
$C_{1}=4\pi\zeta(0)$. Therefore
\begin{equation}
C_{1}=C^{N}_{1}-2\int_{\partial{\cal M}}d^{d-1}x\sqrt{h}\:\psi
\end{equation}
-- a well-known result.

Finally, we find
\begin{eqnarray}
C^{(d)}_{2}\Big|_{\cal I}&=&C^{(d)N}_{2}\Big|_{\cal I}+
\sum_{(ij)}\int_{{\cal I}_{ij}}d^{d-2}x\sqrt{\gamma}\:\Bigg\{ 
\frac{\pi}{4}\bigg[-2\left(\kappa_{i}\psi_{j}+\kappa_{j}\psi_{i}\right)
+\left(\widehat{\kappa}_{i}\psi_{j}+\widehat{\kappa}_{j}\psi_{i}\right)
\nonumber\\
& &+2\left(\psi^{2}_{i}+\psi^{2}_{j}\right)\bigg]
+\frac{1}{3}\bigg[
-5\left(\kappa_{i}\psi_{j}+\kappa_{j}\psi_{i}\right)
+3\left(\widehat{\kappa}_{i}\psi_{i}+\widehat{\kappa}_{j}\psi_{j}\right)
\nonumber\\
& &+12\psi_{i}\psi_{j}
-3\left(\nabla_{\widehat{n}_{i}}\psi_{i}+\nabla_{\widehat{n}_{j}}
\psi_{j}\right)\bigg]\Bigg\}\\
C^{(d)N}_{2}\Big|_{\cal I}&=&
\sum_{(ij)}\int_{{\cal I}_{ij}}d^{d-2}x\sqrt{\gamma}\:
\Bigg\{\frac{\pi}{384}\bigg[
24(1-4\xi)R-12\left(\widehat{R}_{i}+\widehat{R}_{j}\right)+16\widetilde{R}
\nonumber\\
& &+24\left(\kappa^{2}_{i}+\kappa^{2}_{j}\right)
+48\left(\hbox{tr}\kappa^{2}_{i}+\hbox{tr}\kappa^{2}_{j}\right)
+-15\left(\widehat{\kappa}^{2}_{i}+\widehat{\kappa}^{2}_{j}\right)
-42\left(\hbox{tr}\widehat{\kappa}^{2}_{i}+\hbox{tr}\widehat{\kappa}^{2}_{j}
\right)\bigg]\nonumber\\
& &+\frac{1}{270}\bigg[
3R^{\mu\nu}\left(n^{\mu}_{i}\widehat{n}^{\nu}_{i}
+n^{\mu}_{j}\widehat{n}^{\nu}_{j}\right)
-434\left(\kappa_{i}\widehat{\kappa}_{i}
+\kappa_{j}\widehat{\kappa}_{j}\right)
+408\kappa_{i}\kappa_{j}\nonumber\\
& &+490\widehat{\kappa}_{i}\widehat{\kappa}_{j}
+52\hbox{tr}\widehat{\kappa}_{i}\widehat{\kappa}_{j}
+88\left(\nabla_{\widehat{n}_{i}}\kappa_{i}
+\nabla_{\widehat{n}_{j}}\kappa_{j}\right)
\bigg]\Bigg\} 
\end{eqnarray}

\section{Conclusions}
 
We have not been able to obtain  the codimension-2 integrand, ${\cal S}_{ij}$, 
in full for general 
$\theta$, except when all extrinsic curvatures vanish. Relaxing this 
restriction seems to be the most desirable extension of our work, and would 
probably involve a complex problem in specific-case evaluation. 

Another interesting extension 
would be to find ${\cal S}_{ij}$ for general mixed boundary conditions in terms
of the $P_{\pm}$ operators, thus completing the calculations started in 
section 10. As we have noted there, with mixed boundary conditions 
the heat kernel on a product manifold is not in general a product. 
Work on mixed conditions  has been done in \cite{bg} for 
the smooth boundary case.

A problem related to ours is that of conical singularities. These exist on, for
example, a lune with periodic boundary conditions, so that there is a 
codimension-2 submanifold, but not one of codimension-1. Fursaev \cite{fursaev}
has evaluated the heat-kernel coefficients in this situation, and has derived 
expressions similar to the intrinsic part of ours.

The perturbative approach implemented in section 11 may be useful elsewhere in 
the ongoing work of determining heat-kernel coefficients. As an 
alternative to 
perturbing the boundary function, a perturbation in the coordinates of the 
boundary, and hence in $\kappa$, yields a simple expression for the 
first-order effect on the eigenvalues, possibly enabling terms in the $C_{k}$ 
to first-order in $\kappa$ to be calculated. 

\vspace*{5mm}

{\bf Acknowledgement:} The authors would like to thank Klaus Kirsten for some 
very helpful discussions.

\end{document}